\documentclass[iop, aj]{emulateapj}
\usepackage{threeparttable}
\usepackage{multirow}
\usepackage{times}
\usepackage{enumitem}
\usepackage{url}
\usepackage{amsmath,bm}
\usepackage{color,graphicx}

\shorttitle{}
\shortauthors{Y.B. Li et al.}

\begin{document}

\title{A new hyper-runaway star discovered from LAMOST and Gaia: ejected almost in the Galactic rotation direction}

\author{Yin-Bi Li\altaffilmark{1}}
\author{A-Li Luo$^{*}$\altaffilmark{1}}
\author{Gang Zhao$^{*}$\altaffilmark{1}}
\author{You-Jun Lu$^{*}$\altaffilmark{1}}
\author{Xue-Sen Zhang\altaffilmark{2}}
\author{Fu-Peng Zhang\altaffilmark{3}}
\author{Bing Du\altaffilmark{1}}
\author{Fang Zuo\altaffilmark{1}}
\author{Lan Zhang\altaffilmark{1}}
\author{Yang Huang\altaffilmark{4,5}}
\author{Mao-Sheng Xiang\altaffilmark{1,5}}
\author{Jing-Kun Zhao\altaffilmark{1}}
\author{Yong-Heng Zhao\altaffilmark{1}}
\author{Zhan-Wen Han\altaffilmark{6}}

\altaffiltext{1}{Key Laboratory of Optical Astronomy, National Astronomical Observatories, Chinese Academy of Sciences, Beijing 100101, China; {\it lal@bao.ac.cn {\rm (ALL)}; gzhao@bao.ac.cn {\rm (GZ)}; luyj@bao.ac.cn {\rm (YJL)}}}
\altaffiltext{2}{ExtantFuture (Beijing)Technology Co.,Ltd, Beijing 100102, China;}
\altaffiltext{3}{School of physics and astronomy, Sun Yat-sen University, Guangzhou 510275, china;}
\altaffiltext{4}{South-Western Institute for Astronomy Research, Yunnan University, Kunming 650500, China;}
\altaffiltext{5}{LAMOST Fellow}
\altaffiltext{6}{Key Laboratory for the Structure and Evolution of Celestial Objects, Yunnan Observatories,Chinese Academy of Sciences, Kunming 650216, China;}

\begin{abstract}
In this paper, we report the discovery of a new late-B type unbound hyper-runaway star (LAMOST-HVS4) from the LAMOST spectroscopic surveys. According to its atmospheric parameters, it is either a B-type main-sequence (MS) star or a blue horizontal branch (BHB) star. Its Galactocentric distance and velocity are 30.3 $\pm$ 1.6~kpc and 586$\pm$7 km s$^{-1}$ if it is an MS star, and they are 13.2 $\pm$ 3.7~kpc and 590$\pm$7 km s$^{-1}$ if a BHB star. We track its trajectories back, and find that the trajectories intersect with the Galactic disk and the Galactic center lies outside of the intersection region at the 3$\sigma$ confidence level. We investigate a number of mechanisms that could be responsible for the ejection of the star, and find that it is probably ejected from the Galactic disk by supernova explosion or multiple-body interactions in dense young stellar clusters.

\end{abstract}

\section{Introduction}

Hypervelocity stars (HVSs) are moving so fast that they may escape from the Galaxy. A natural explanation is that they were ejected out from the Galactic center (GC)
by interactions between stars and the massive black hole (MBH) or hypothetical (binary) MBHs, as predicted by \cite{1988Natur.331..687H} and \cite{2003ApJ...599.1129Y}.
In the Galactic center origin scenario, three different mechanisms can produce HVSs, i.e., (1) tidal breakup of binary stars by a single MBH
\citep{1988Natur.331..687H, 2003ApJ...599.1129Y, 2006ApJ...653.1194B}, and the binary stars could be injected into the vicinity of the MBH from the young stellar disk in the GC
\citep[e.g., ][]{2010ApJ...709.1356L, 2010ApJ...722.1744Z, 2013ApJ...768..153Z} or from the Galactic bulge \citep{2009ApJ...690..795P, 2009ApJ...698.1330P}; (2) interactions between single stars and a hypothetic binary MBH \citep{2003ApJ...599.1129Y, 2006ApJ...651..392S, 2007MNRAS.379L..45S, 2006ApJ...648..976M}; (3) interactions between single stars and a cluster of stellar mass black holes around the MBH \citep{2008MNRAS.383...86O}. Aside from the GC origins, alternative mechanisms were also proposed to explain HVSs, including the tidal debris of an accreted and disrupted dwarf galaxy \citep{2009ApJ...691L..63A, 2009ApJ...707L..22T}, the surviving companion stars of Type Ia supernova (SN Ia) explosions \citep{2009A&A...508L..27W}, the result of dynamical interactions between multiple stars (e.g, Gvaramadze et al. 2009), and the runaways ejected from the Large Magellanic Cloud (LMC), assuming that the latter hosts an intermediate mass black hole \citep{2016ApJ...825L...6B, 2017MNRAS.469.2151B}.

Before Gaia DR2 release, there are 38 hyper velocity star with unbound probabilities larger than 0.5 discovered \citep{2005ApJ...622L..33B, 2005A&A...444L..61H, 2005ApJ...634L.181E, 2006ApJ...647..303B, 2008A&A...483L..21H, 2009ApJ...690.1639B, 2009A&A...507L..37T, 2010ApJ...711..138I, 2012ApJ...744L..24L, 2012ApJ...751...55B, 2014ApJ...787...89B, 2014ApJ...785L..23Z, 2015RAA....15.1364L, 2017ApJ...847L...9H, 2017Sci...357..680V, 2018Boubert}. Most of them are massive B-type stars in the Galactic halo, and are consistent with the GC origin for ejecting unbound HVSs \citep{2015ARA&A..53...15B}. The six exceptions are US\,708 (HVS2), HE~0437-5439(HVS3), HD~271791, SDSS~J013655.91+242546.0, LAMOST J115209.12+120258.0 (Li10), and LP~40-365. US\,708 is a helium-rich sub-dwarf O-type star with a mass of $\sim$\,0.3\,M$_{\odot}$, and LP~40-365 is a low mass white dwarf (WD). Their trajectories favor an origin of tidal interactions in binaries, and they are probably remnants of SN Ia explosions \citep{2015Sci...347.1126G, 2017Sci...357..680V}, originating from the WD+He star systems (see Wang et al. 2009, 2013). HE~0437-5439 is a B-type star, and is likely to be originated from the centre of the Large Magellanic Cloud (LMC) \citep{2018Erkal}. Other three exceptions are all hyper-runaway stars ejected from the Galactic disk, and they are B- (HD~271791), A- (SDSS~J013655.91+242546.0), or F-type (LAMOST J115209.12+120258.0) stars, respectively \citep{2008A&A...483L..21H, 2009A&A...507L..37T, 2015RAA....15.1364L}.

Using Gaia DR2, a total of 58 high velocity star candidates were discovered from a subset with radial velocity measurements \citep{2018Marchetti, 2018Hattori}. Among them, there are 9 hyper-runaway star candidates. In addition, three new hyper-velocity WDs are found by \citet{2018Shen}, and they were previously the WD companions to primary WDs that exploded as SNe Ia.

In this paper, we continue to search for massive hyper-velocity stars from LAMOST surveys. The paper is organized as follows. We provide a brief description of the LAMOST data in Sub-section 2.1, and we focus on estimating the Galactocentric radial velocity, atmospheric parameters, absolute magnitude, and heliocentric distance of the LAMOST-HVS4 in Sub-section 2.2. We then investigate whether it is unbound to the Galaxy with five potential models of the Galaxy in Sub-section 2.3, and calculate its space position and velocity in Sub-section 2.4. Its possible origin is discussed in Section\,3, and conclusion is summarized in Section\,4.

\section{The LAMOST survey and the new HVS}
\subsection{Data}
LAMOST is a 4 meter quasi-meridian reflecting Schmidt telescope, which is equipped with 4000 fibers and can simultaneously obtain 4000 ($\lambda\lambda$ 3700 -- 9000 $\rm\AA$) low resolution ($R$\,$\sim$\,$1800$) spectra per exposure \citep{1996ApOpt..35.5155W, 2004ChJAA...4....1S, 2012RAA....12..723Z, 2012RAA....12.1243L, 2012RAA....12.1197C}.
After a one year pilot survey starting in 2011, the five-year LAMOST Regular Surveys were initiated in the fall of 2012 and completed in the summer of 2017. LAMOST Phase-II includes low and medium resolution surveys, and the low resolution survey began in January 2018, which is the continuation of Phase-I. According to the LAMOST data policy, the Phase-II surveys released data productions of the first three months to domestic users and their internal co-workers (DR6$-$QI) in January 2018, which include 203,200 low-resolution spectra data.

We performed a systematic search for HVSs in the DR6-QI data set, and we find that one star turned out to be an HVS (hereafter, denoted by LAMOST-HVS4).

\subsection{Properties of LAMOST-HVS4}

%rv and pm
The star (J225837.56+400005.2; LAMOST-HVS4) was observed on November 21, 2017, and it has magnitudes (g/r/i/z/y) of the Panoramic
Survey Telescope and Rapid Response System (Pan-STARRS; Kaiser et al. 2002, 2010; Chambers et al. 2016) larger than 16.7 mag.
The heliocentric radial velocity is $v_{\rm r\odot}=359 \pm 7$ km s$^{-1}$, and we check it using the cross-correlation package RVSAO of IRAF \citep{1998PASP..110..934K}. The value is  consistent with the radial velocity provided by LAMOST. We translate the heliocentric radial velocity ($v_{\rm r\odot}$) to a Galactocentric radial component of $v_{\rm rf}=585 \pm 7$ km s$^{-1}$ according to

\begin{equation}
v_{\rm rf}=v_{\rm r\odot} + U_0\cos l \cos b + (V_{\rm LSR}+V_0)\sin l \cos b + W_0\sin b,
\end{equation} \\
where we adopt $V_{\rm LSR} = 235$ km s$^{-1}$ for the motion of the local standard of rest (LSR) \citep{2005ApJ...629..268H, 2012ApJ...759..131B, 2014ApJ...783..130R} and $(U_0, V_0, W_0)=(9.58, 10.52, 7.01)$ km s$^{-1}$ for the peculiar motion of the Sun with respect to the LSR (Tian et al. 2015; c.f., Sch\"{o}nrich et al. 2010; Huang et al. 2015).

\begin{figure*}
\begin{center}
\includegraphics[scale=0.45,angle=0]{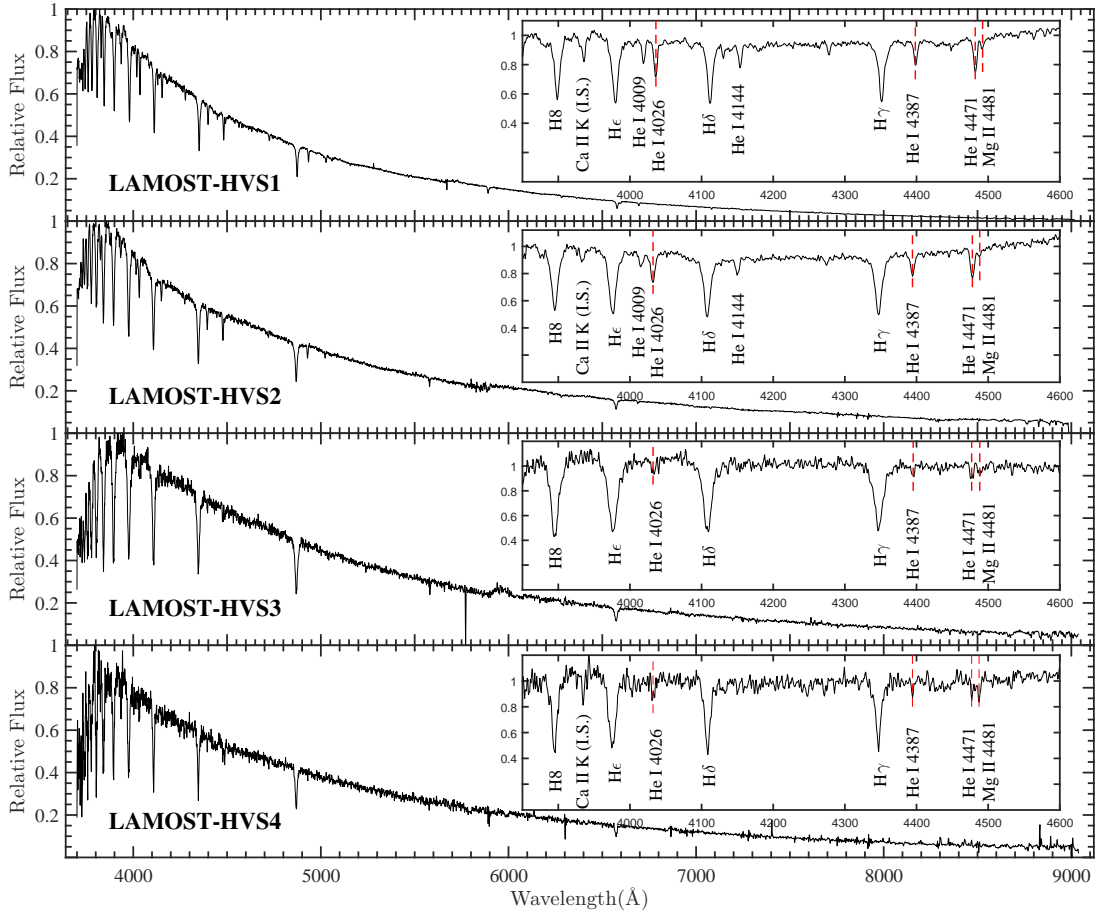}
\caption{The LAMOST spectra of LAMOST-HVS1-4 (from top to bottom). The inset in each panel shows the enlarged normalized spectrum from 3850 $\rm\AA$ to 4600 $\rm\AA$ for a better view of the spectral lines. \label{fig:spectra}}
\end{center}
\end{figure*}

The spectrum of LAMOST-HVS4 is shown at the bottom of Figure~\ref{fig:spectra}, and the spectra of three previous LAMOST HVSs are also displayed in this figure for comparison. The inset in each panel shows the enlarged normalized spectrum from 3850 to 4600 $\rm\AA$ for a better view, and the four red dashed lines in it mark He I and Mg II lines. Similar to ``LAMOST-HVS3'', the He I lines at 4026 $\rm\AA$, 4387 $\rm\AA$, and 4471 $\rm\AA$ and the Mg II line at 4481 $\rm\AA$ are marginally detected, which indicates that LAMOST-HVS4 is a late B-type star. Its basic atmospheric parameters (effective temperature $T_{\rm eff}$, surface gravity $\log[g]$, metallicity $[{\rm Fe/H}]$), and heliocentric radial velocity ($v_{\rm r\odot}$) are calculated from the LAMOST spectrum using the e LAMOST stellar parameter pipeline (LASP; Luo et al.(2015)). The results show that $T_{\rm eff} = 15140 \pm 578~{\rm K}$, $\log[g] = 3.9 \pm 0.3$, $[{\rm Fe/H}] = 0.29 \pm 0.18$, and $v_{\rm r\odot} = 359$ km $s^{-1}$. However, classification by $T_{\rm eff}$ and $\log[g]$ is ambiguous, and we cannot reliably distinguish between a B-type main-sequence star (MS) and a hot blue horizontal branch (BHB) star based on its $T_{\rm eff}$ and $\log[g]$. Although, the metallicity of LAMOST-HVS4 is roughly solar, which suggests it is probably a B type MS star, we also discuss the case that it is a BHB star.

We obtain parallax and proper motions of LAMOST-HVS4 from the Gaia DR2 catalog. Unfortunately, the parallax of LAMOST-HVS4 is negative (-0.113 mas), which suggests it is not a nearby star with a heliocentric distance lower than 3 or 5 kpc and no valid heliocentric distance is available from the Gaia DR2 parallax (private communication with Bailer-Jones). Since the nature and heliocentric distance for LAMOST-HVS4 cannot be inferred from its Gaia parallax, we estimate the distances using the photometric distance method for two cases: (1) LAMOST-HVS4 is an MS star; (2) it is a BHB star in this paper.

If it is a B-type MS star, its absolute magnitude, mass, and lifetime can be inferred by using the look-up table provided by Mamajek \footnote{\url{http://www.pas.rochester.edu/$\sim$emamajek/EEM\_dwarf\_UBVIJHK\_colors\newline\ Teff.txt}} \citep{2017ApJ...847L...9H}. We find it is a B6 type star with $M_{V}(\rm MS) = -0.69 \pm 0.12$~mag, $\rm Mass \approx 4.3~\rm M_{\sun}$, and $\rm Age \approx 55~\rm Myr$, consistent with the spectral type given by the LAMOST 1D pipeline \citep{2015RAA....15.1095L}. If it is a BHB star, its absolute magnitude is estimated to be $M_{V}(\rm BHB) = 1.72 \pm 0.88$ mag using the relationship of \citet{2002MNRAS.337...87C}, and the lifetime of such a BHB star is roughly 13 Gyr \citep{1996ApJ...466..359D, 2013A&A...549A.145L, 2016MNRAS.463.3449L}.

We cross-match with photometry catalogs of the Sloan Digital Sky Survey (SDSS; Alam et al. 2015), Pan-STARRS, Two Micron All Sky Survey (2MASS; Skrutskie et al. 2006), and the AAVSO Photometric All Sky Survey(APASS; Henden et al. 2016) with a search radius of 5 arcsec, and obtain magnitudes from the Pan-STARRS catalog as listed in Table \ref{tab:tab1}. For the SDSS, 2MASS or APASS catalog, no target is found within the radius of 5 arcsec, and we note that LAMOST-HVS4 is outside of the SDSS footprint. The Johnson $B$ and $V$ magnitudes in this table are converted from the $g$ and $r$ magnitudes of Pan-STARRS survey using the transformation relationship presented in \citet{2012ApJ...750...99T}. With the apparent and absolute magnitudes derived, we estimate the heliocentric distance after applying a reddening correction of \citet{2011ApJ...737..103S}. The heliocentric distance of $d_{\rm MS} = 27.9 \pm 1.5~\rm kpc$ if it is a B-type MS star, and it is $d_{\rm BHB} = 9.2 \pm 3.7~\rm kpc$ if it is a  hot BHB. Thus, we consider 9.2 kpc to be the lower limit of the heliocentric distance for the LAMOST-HVS4.

\begin{table*}
\centering
\caption{Properties of the LAMOST-HVS4. \label{tab:tab1}}
\begin{threeparttable}
\begin{tabular}{ll}
\hline
 & LAMOST-HVS4\\
\hline

{Position ($J2000$)}  &  ($\alpha$, $\delta$) = (344\fdg65650, 40\fdg001470) \\
                      &  ($l$, $b$) = (100\fdg568290, -17\fdg920979) \\
Magnitudes (mag)    & $g = 16.79 \pm 0.01$ \tnote{a} \\
                    & $r = 16.96 \pm 0.004$ \tnote{a} \\
                    & $i = 17.18 \pm 0.005$ \tnote{a} \\
                    & $z = 17.30 \pm 0.005$ \tnote{a} \\
especially                    & $y = 17.36 \pm 0.01$ \tnote{a} \\
                    & $B = 16.90 \pm 0.04$ \tnote{b} \\
                    & $V =  16.89 \pm 0.02$ \tnote{b}\\

$E(B-V)$    & 0.1129 \tnote{c} \\

$S/N\_g$, $S/N\_r$, $S/N\_i$ & 22, 21, 22 \\

Spectral type & B6V \\

$T_{\rm eff}$ (K) & $15140 \pm 578$\\

log$g$ (dex) & $3.9 \pm 0.30$ \\

$[Fe/H]$ (dex) & $0.29 \pm 0.18$ \\

Mass ($M_{\odot}$) & $\sim$ 4.3 (MS) \\

lifetime (Myr)  & $\sim$\ 55 $\pm$ 6 \\

Absolute magnitude & $M_{\rm V} = -0.69 \pm 0.12$ (MS) \\
                   & $M_{\rm V} = 1.72 \pm 0.88$ (BHB) \\

Heliocentric distance (kpc)\tnote{c} & 27.9 $ \pm $ 1.5 (MS) \\
                                     & 9.2  $ \pm $ 3.7 (BHB) \\

Galactocentric distance (kpc) & 30.3 $ \pm $ 1.6 (MS) \\
                              & 13.2 $ \pm $ 3.7 (BHB) \\

Radial velocity (km\,s$^{-1}$) &  $v_{r\odot} = 359 \pm 7$ \\
                               &  $v_{\rm rf}$ = 585 $\pm 7$ \\

Proper Motion (mas yr$^{-1}$) & $(\mu_\alpha\cos\delta,\, \mu_\delta) = $ \\
                              & (-3.9 $\pm$ 5.4, 1.6 $\pm$ 5.4) [PPMXL] \\
                              & (-1.644 $\pm$ 2.431, 1.579 $\pm$ 2.431) [HSOY] \\
                              & (4.685$\pm$2.174, -1.137$\pm$1.441) [Tian+2017] \\
                              & (0.343$\pm$0.093, -0.288$\pm$0.101) [GAIA DR2] \\
                              %& (-3.3803$\pm$1.5521, -0.3291$\pm$1.2083) [Weighted Average of three catalogs] \\
                              %& (-3.3336$\pm$1.6205, -0.4308$\pm$1.2396) [Weighted Average of HSOY and Tian+2017 catalogs] \\
Flight time (Myr) & 48$\pm$3 (MS) \\
                  & 20$\pm$7 (BHB) \\
\hline
\end{tabular}
\begin{tablenotes}
\item[a] The $g$, $r$, $i$, $z$, and $y$ magnitudes of the Pan-STARRS survey.
\item[b] The Johnson $B$ and $V$ magnitudes converted from the $g$ and $r$ magnitudes measured by the Pan-STARRS survey.
\item[c] $E(B-V)$ is estimated with the method mentioned in \citet{2011ApJ...737..103S}
\end{tablenotes}
\end{threeparttable}
\end{table*}

\subsection{Unbound to the Galaxy ?}
In this sub-section, five Galactic potential models \citep{1990ApJ...348..485P, 2005ApJ...634..344G, 2008ApJ...684.1143X, 2010ApJ...712..260K, 2014ApJ...793..122K} are adopted to calculate escape velocities ($v_{\rm esc}$) at different Galactocentric distances ($r$), and to further determine whether LAMOST-HVS4 is unbound to the Galaxy. Hereafter, we abbreviate the five models as Paczynski+1990, Gnedin+2005, Xue+2008, Koposov+2010, and Kenyon+2014, respectively. Among them, the Paczynski+1990, Koposov+2010, and Kenyon+2014 are divergent models, while the Gnedin+2005 and Xue+2008 are convergent models. For the two convergent models, we calculate $v_{\rm esc}$ at different $r$ using $v_{\rm esc} = \sqrt{2|\phi(r) - \phi(\infty)|}$, where $\phi(r)$ and $\phi(\infty)$ are Galactic potentials at distances of r and infinity. For three divergent models, we estimate $v_{\rm esc}$ according to $v_{\rm esc} = \sqrt{2(|\phi(r) - \phi(200~\rm kpc)|) + (200)^{2}}$ \citep{2008ApJ...680..312K}, and the number ``200'' in the second term represents minimum value of velocity ($=$ 200 km s$^{-1}$) for a unbound star at $r = 200$ kpc.

\begin{figure*}
\begin{center}
\includegraphics[scale=0.3,angle=0]{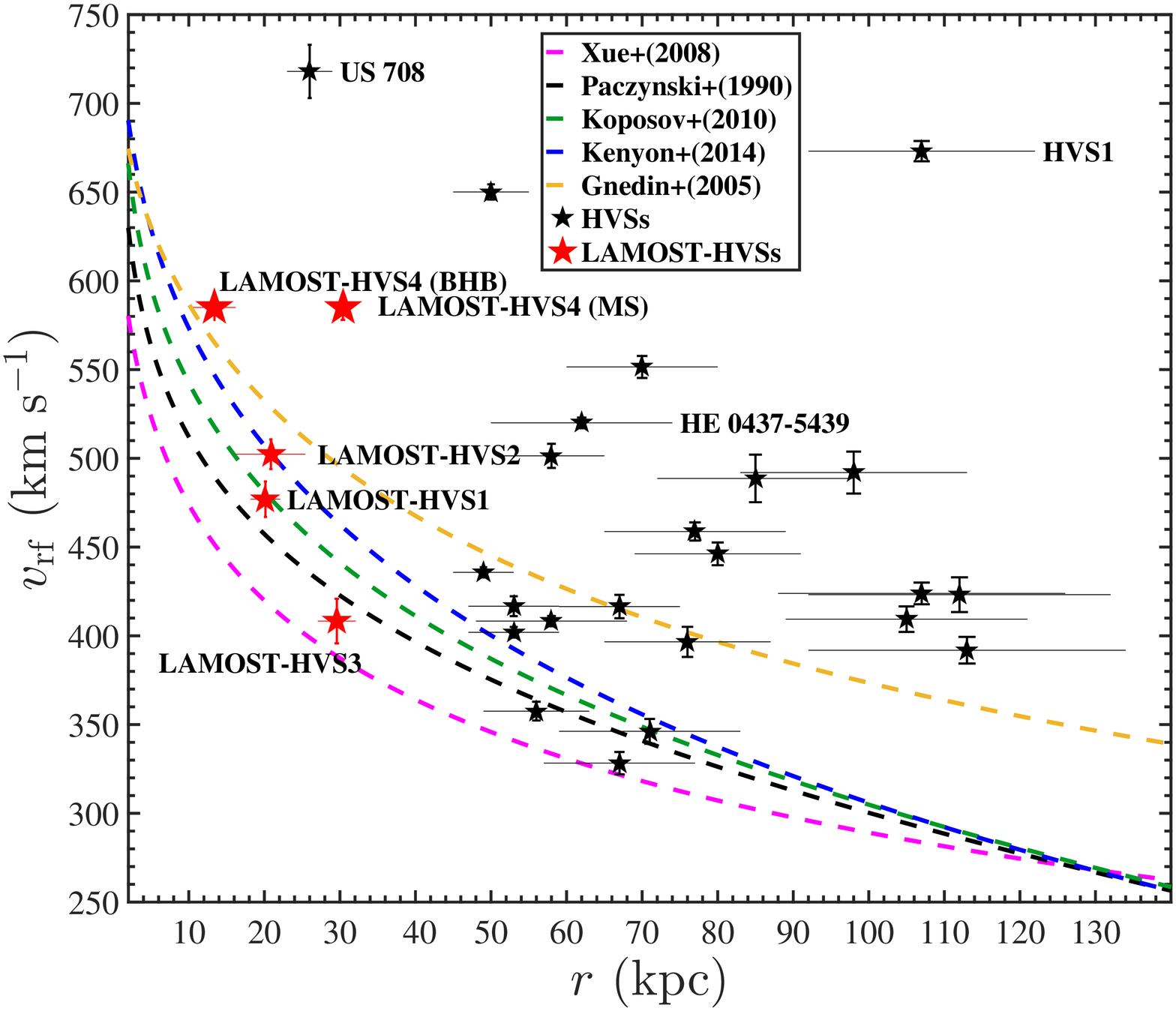}
\caption{Galactocentric radial velocities of known HVSs (Edelmann et al. 2005; Hirsch et al. 2005; Brown et al. 2014; Zheng et al. 2014; Huang et al. 2017; ) and LAMOST-HVS4 versus the Galactocentric distances. The five short dashed curves are escape velocities determined from the five Galactic potential models (Paczynski+1990; Gnedin+2005; Xue+2008; Koposov+2010; Kenyon+2014), and the difference illustrates the current uncertainties between these models. There are two larger red filled pentagrams, which represent the LAMOST-HVS4 as a B-type MS star or a BHB, respectively. The three smaller red filled pentagrams are the other three LAMOST HVSs (Zheng, et al. 2014; Huang et al. 2017). Other black filled pentagrams represent 23 identified B-type HVSs (Edelmann et al. 2005; Hirsch et al. 2005; Brown et al. 2014;) \label{fig:potential}}
\end{center}
\end{figure*}

With the determined velocities ($v_{\rm rf}$) and distances ($r$), we show LAMOST-HVS1-4 (red filled pentagrams) and another 23 identified B-type HVSs (black filled pentagrams) in Figure~\ref{fig:potential}. We also plot $v_{\rm esc}$ at different $r$ with color short-dashed lines based on the above potential models. As seen from this figure \footnote{LAMOST-HVS1 is unbound under the Xue+2008 and Paczynski+1990 models, LAMOST-HVS2 is unbound under the Xue+2008, Koposov+2010 and Paczynski+1990 models, and LAMOST-HVS3 is unbound under the Xue+2008 model. \citet{2017ApJ...847L...9H} mentioned the three LAMOST HVSs are all unbound to the Milky Way under their Milky Way mass model \citep{2016MNRAS.463.2623H, 2017submitted}. Technically, the definition of ``bound'' or ``unbound'' depends on the choice of the Galaxy mass model especial for those stars in the real case marginally bound or unbound to the Galaxy.}, the velocity $v_{\rm rf} = 585$~km s$^{-1}$ of LAMOST-HVS4 is above the escape velocities obtained from all the potential models at $R = 13.4 \pm 2.8~\rm kpc$ and $R = 30.4 \pm 1.4~\rm kpc$, which represents LAMOST-HVS4 is an unbound HVS whether it is an MS or a BHB star.

\subsection{Space position and velocity}

In this sub-section, we determine the phase-space coordinate for LAMOST-HVS4 with heliocentric distance, radial velocity, proper motions, equatorial and galactic coordinates. The right hand Cartesian coordinate system adopted here is centered on the GC: the X-axis points from the Sun to the GC with the Sun at x = -8 kpc; the Y-axis points in
the direction of Galactic rotation; the Z-axis points toward the Northern Galactic Pole.

Proper motions of LAMOST-HVS4 are obtained using the ``VizieR'' tool of SIMBAD database \footnote{http://vizier.u-strasbg.fr/viz-bin/VizieR}. It has a proper motion of $(\mu_\alpha\cos\delta,\, \mu_\delta) = (-3.9\pm5.4,\, 1.6\pm 5.4)\, {\rm mas\, yr^{-1}}$ in PPMXL catalog \citep{2010AJ....139.2440R}, $(-1.644\pm2.431,\, 1.579\pm2.431)\, {\rm mas\, yr^{-1}}$ in HSOY catalog \citep{2017A&A...600L...4A}, $(-4.685\pm2.174,\, -1.137\pm1.441)\, {\rm mas\, yr^{-1}}$ in the catalog of \citet{2017ApJS..232....4T} (hereafter Tian+2017), and $(0.343\pm0.093,\, -0.288\pm0.101)\, {\rm mas\, yr^{-1}}$ in Gaia DR2 catalog \citep{2018A&A1804}, respectively. These measurements are listed in Table~\ref{tab:tab1}, and the Gaia DR2 measurements lie within the error of PPMXL and HSOY proper motions but have significantly higher accuracy. Thus, we adopt the proper motion from Gaia DR2 in this paper.

Using Monte-carlo method, we obtain the phase space coordinates of LAMOST-HVS4 as listed in Table~\ref{tab:tab2}, and its Galactocentric distance (r) and total space velocity ($V_{\rm G}$) are also displayed in the table. From this table, we note that LAMOST-HVS4 locates 30.4 kpc away from the GC when it is an MS star, and 13.4 kpc from the GC when it is a BHB star. In addition, it lies at 8.6 kpc or 2.8 kpc below the Galactic disk respectively.

\begin{table*}
\centering
\caption{Space positions and velocities. \label{tab:tab2}}
\begin{threeparttable}
\begin{tabular}{ccc}
\hline
 & ~~~~MS & ~~BHB \\
($x$, $y$, $z$)~\tnote{a}~~~(kpc) & (-12.9$\pm$0.3, 26.1$\pm$1.4, -8.6$\pm$0.5) & (-9.6$\pm$0.6, 8.6$\pm$3.4, -2.8$\pm$1.1)  \\
r~\tnote{b}~~~(kpc) & 30.4$\pm$1.4 & 13.4$\pm$2.8 \\
($v_{\rm x}$, $v_{\rm y}$, $v_{\rm z}$)\tnote{c}~~~(km s$^{-1}$) & (-74$\pm$12, 560$\pm$8, -155$\pm$13) & (-60$\pm$5, 574$\pm$7, -121$\pm$8)\\
$V_{\rm G}$\tnote{d}~~~(km s$^{-1}$) & 586$\pm$7 & 590$\pm$7 \\
\hline
\end{tabular}
\begin{tablenotes}
\item[a] Space positions ($x$, $y$, $z$) when it is an MS star or a BHB star.
\item[b] Galactocentric distance when it is an MS star or a BHB star.
\item[c] Space velocities ($v_{\rm x}$, $v_{\rm y}$, $v_{\rm z}$) when it is an MS star or a BHB star.
\item[d] Space total velocities when it is an MS star or a BHB star.
\end{tablenotes}
\end{threeparttable}
\end{table*}

\section{Possible Origin}

In this section, we trace the trajectory of LAMOST-HVS4 back, and investigate its possible origins.

We perform two Monte-Carlo simulations using cloud computation on Alibaba cloud\footnote{http://astrocloud.china-vo.org}, by assuming that the measurement errors of position and velocity follow Gaussian distributions with the standard deviations listed in Table~\ref{tab:tab2}. The first simulation is performed assuming that LAMOST-HVS4 is a B-type MS star, and the second simulation assumes that it is a BHB star. Each simulation adopts the five Galactic potential models (Paczynski+1990; Gnedin+2005; Xue+2008; Koposov+2010; Kenyon+2014) as mentioned above, and 50,000 trajectories are calculated for each model. Since the lifetime of LAMOST-HVS4 is about 55 Myr for the case of an MS star or 13 Gyr for the BHB case, we trace trajectories back to 55 Myr ago in the first simulation, and 13 Gyr ago in the second one.

\begin{figure*}
\begin{center}
\includegraphics[scale=0.3,angle=0]{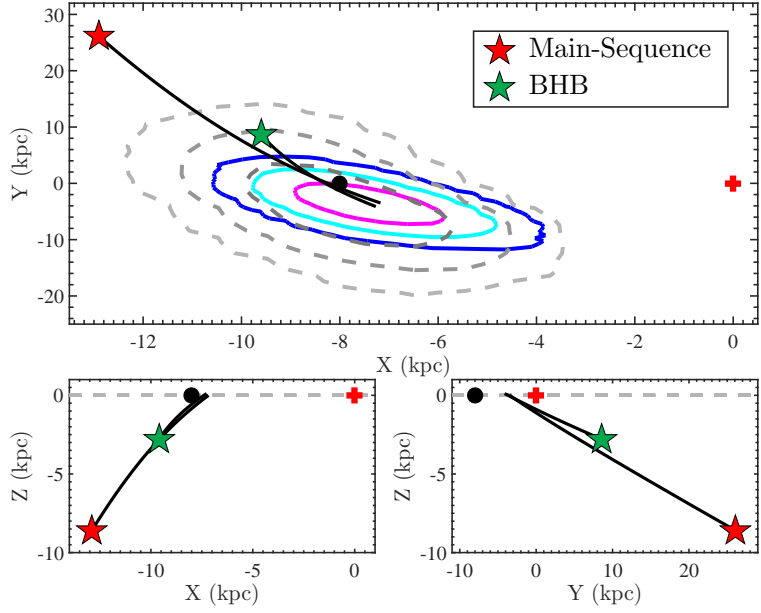}
\caption{\textbf{Two-dimensional projections of the orbit of LAMOST-HVS4 under the Kenyon+2014 potential model in the Galactic rectangular coordinates. Symbols mark the present positions of LAMOST-HVS4 (red and green pentagram), the GC (red filled plus), and the Sun (black filled circle). The color contours mark the 1/2/3-$\sigma$ confidence levels of intersection points between the trajectories and the Galactic disk if it is an MS star, and the gray dashed contours show intersection regions if it is a BHB star.}\label{fig:origin_place}}
\end{center}
\end{figure*}

All trajectories intersect with the Galactic disk in the first simulation, and approximately 99.42$\%$ of the trajectories cross the disk in the second simulation. Thus, we calculate intersections of these trajectories with the Galactic disk for each potential model, and we obtain their 1/2/3-$\sigma$ confidence regions. The results show that intersection regions in the different potential models are extremely close, and we only display the results of the Kenyon+2014 model in Figure~\ref{fig:origin_place}.

In this figure, the sun and GC are marked by a black filled circle and a red filled plus, respectively, and the current position of LAMOST-HVS4 is displayed with pentagrams. The red pentagram represents the current position if LAMOST-HVS4 is an MS star, and the green pentagram shows the present position if it is a BHB star. Colored contours represent its origin on the Galactic disk. The magenta, cyan and blue contours show the origin places corresponding to 1, 2, and 3 $\sigma$ confidence levels if LAMOST-HVS4 is an MS star, and three dashed gray contours show origin areas of 1, 2, and 3 $\sigma$ level if it is a BHB star. From this figure, we note that LAMOST-HVS4 was originated from the regions near the sun, and the GC origin is excluded up at a 3 $\sigma$ confidence level.

The flight time from the current position to the intersection points on the Galactic disk is about 48$\pm$3 Myr if it is an MS star and 20$\pm$7 Myr if it is a BHB star. This indicates that it can move to its present position if it is originated from the Galactic disk, even though the lifetime ($\sim$ 55 Myr) is not much longer than the flight time when it is an MS. In addition, the ejected velocities from the Galactic disk under the different potential models are nearly the same, and it is 697$\pm$12 km s$^{-1}$ or 623$\pm$25 km s$^{-1}$ if it is an MS star or a BHB star. As mentioned in \citet{2015ARA&A..53...15B}, hyper-runaway stars in sun circle ejected at the rotation direction require ejection velocities $\geq$ 400 km s$^{-1}$, thus ejected velocity of LAMOST-HVS4 satisfies the requirement. We further estimate the angles between the ejected velocities of LAMOST-HVS4 from the disk and the velocity of Galactic disk, and it is 27.9$^{\circ} \pm$ 1.8$^{\circ}$ or 23.1$^{\circ} \pm$ 4.0$^{\circ}$ for the MS or BHB case. Thus, we can conclude that LAMOST-HVS4 is probably an unbound hyper-runaway star originated from the Galactic disk, and was ejected almost along the Galactic disk rotation direction.

There are two main dynamic mechanisms to produce hyper-runaway stars: 1) supernova explosion in stellar binary systems \citep{1961BAN....15..265B, 2000ApJ...544..437P}; 2) dynamic encounters between stars in dense stellar systems such as young stellar clusters \citep{1967BOTT....4...86P, 1990AJ.....99..608L, 2009MNRAS.396..570G}. We thus search for H\,{\small \bf II} regions, i.e., stellar formation regions, in its origin places from a catalog of H\,{\small \bf II} regions \citep{2014ApJS..212....1A}, and find that there are at least 21, 166, and 244 H\,{\small \bf II} regions in the origin area of 1, 2, and 3 $\sigma$ confidence level if it is an MS star, respectively, and 142, 284, 392 H\,{\small \bf II} regions in the 1, 2, and 3 $\sigma$ origin places if it is a BHB star, respectively. All of these H\,{\small \bf II} regions have estimation of heliocentric distances and lie within $|z| \leq 90~\rm pc$ \citep{2018submitX}. Since the lifetimes of H\,{\small \bf II} regions are about a few million years in general, LAMOST-HVS4 cannot be originated from H\,{\small \bf II} regions above, but it is likely to be ejected from previous H\,{\small \bf II} regions in the intersection area. Combining its ejection angles, LAMOST-HVS4 is extremely likely a hyper-runaway star originated from the star-forming regions by supernova explosions or dynamic 3- or 4-body interactions, and it is the first hyper-runaway star nearly ejected in the direction of Galactic disk rotation. As mentioned in \citet{2015ARA&A..53...15B}, the ejection rate of hyper-runaway stars ejected at the rotation direction is approximately 1$\%$ of the HVS ejection rate, thus they are more rare objects and contribute a small fraction of HVSs.

Tidal disruptions of Milky Way dwarf galaxies can also produce high-velocity stars, as proposed by \citet{2009ApJ...691L..63A}. We calculate intersections between the trajectories of LAMOST-HVS4 and 54 dwarf galaxies within 300 kpc, and its trajectories only cross the LMC and Sagittarius Dwarf Spheroidal Galaxy (Sgr dSph) within a factor 30 times larger than the effective radii. Thus, it is obviously impossible that LAMOST-HVS4 was ejected from the other 52 dwarf galaxies. For LMC and Sgr dSph, we further calculate the nearest distances and flight times of 50,000 orbits. If LAMOST-HVS4 is an MS star, the nearest distances to the LMC and Sgr dSph are 37.9$\pm$0.6 and 24.1$\pm$0.7 kpc, and the flight times are 90$\pm$3 and 48$\pm$2 Myr. The flight time of 90$\pm$3 Myr is larger than its lifetime, thus it is impossible for it to arrive at the LMC within its lifetime. If it is a BHB star, the nearest distances to LMC and Sgr dSph are 35.5$\pm$0.8 and 24.4$\pm$0.8 kpc, and the flight times are 73$\pm$6 and 17$\pm$7 Myr. The tidal radii of LMC and Sgr dSph are 22.3$\pm$5.2 and 3 kpc \citep{2014ApJ...781..121V, 2001MNRAS.323..529H}, respectively, thus LAMOST-HVS4 is unlikely originated from the core regions of LMC and Sgr dSph, but the dwarf galaxy origin cannot be exclude completely.

\section{Summary}
In this paper, we report a new late-B type unbound hyper-runaway star (LAMOST-HVS4) discovered from the LAMOST spectroscopic surveys. Its atmospheric parameters are estimated with the LAMOST spectrum, and it may be a B-type main-sequence (MS) star or a blue horizontal branch (BHB) star. Combing heliocentric distance, position, radial velocity from LAMOST, and proper motions from Gaia DR2, its Galactocentric distance and velocity are 30.3 $\pm$ 1.6~kpc and 586$\pm$7 km s$^{-1}$ if it is an MS star, and they are 13.2 $\pm$ 3.7~kpc and 590$\pm$7 km s$^{-1}$ if it is a BHB star. We integrate its trajectories back, and find that the trajectories intersect with the Galactic disk and the Galactic center is excluded from the intersection region at 3$\sigma$ level. We find that it is probably a hyper-runaway star produced by supernova explosion or multiple-body interactions in dense young stellar clusters, and it was almost ejected at a direction close to that of Galactic rotation.

\section*{Acknowledgements}
We would like to thank the referee for his/her helpful comments.
We thank Shi J.R., Zheng Z., Bailer-Jones C.A.L., Brown Warren R., Gnedin Oleg Y., Kenyon Scott J., Wang Y.F., Wang R., Qin L., Lei Z.X., Li J., Wang J.L., and Xue X.X. for useful discussions.
This work is supported by the National Basic Research Program of China (973 Program, 2014CB845700), the National Natural Science Foundation of China (grant Nos. 11390371/2/4), and L. Z. acknowledges supports from the National Science Foundation of China (NSFC) grants 11773033. This work is supported by the Astronomical Big Data Joint Research Center, co-founded by the National Astronomical Observatories, Chinese Academy of Sciences and the Alibaba Cloud. Guoshoujing Telescope (the Large Sky Area Multi-Object Fiber Spectroscopic Telescope, LAMOST) is a National Major Scientific Project built by the Chinese Academy of Sciences. Funding for the project has been provided by the National Development and Reform Commission. LAMOST is operated and managed by the National Astronomical Observatories, Chinese Academy of Sciences.

\end{document}